\documentclass[physrev,reprint,amsmath,amssymb,superscriptaddress,longbibliography]{revtex4-2}

\usepackage{graphicx}
\usepackage{dcolumn}
\usepackage{bm}
\usepackage{hyperref}
\usepackage{breakurl}
\usepackage{multirow}
\usepackage{enumitem}
\usepackage{color}

\begin{document}

\title{Manipulating Giant Rashba Valley Splitting and Quantum Hall States in Few-Layer Black Arsenic by Electrostatic Gating}

\author{Feng Sheng}
\thanks{Equal contributions}
\affiliation{Zhejiang Province Key Laboratory of Quantum Technology and Device, Department of Physics, Zhejiang University, Hangzhou 310027, P. R. China}

\author{Chenqiang Hua}
\thanks{Equal contributions}
\affiliation{Zhejiang Province Key Laboratory of Quantum Technology and Device, Department of Physics, Zhejiang University, Hangzhou 310027, P. R. China}

\author{Xikang Sun}
\affiliation{Zhejiang Province Key Laboratory of Quantum Technology and Device, Department of Physics, Zhejiang University, Hangzhou 310027, P. R. China}

\author{Jie Hu}
\affiliation{Zhejiang Province Key Laboratory of Quantum Technology and Device, Department of Physics, Zhejiang University, Hangzhou 310027, P. R. China}

\author{Qian Tao}
\affiliation{Zhejiang Province Key Laboratory of Quantum Technology and Device, Department of Physics, Zhejiang University, Hangzhou 310027, P. R. China}

\author{Hengzhe Lu}
\affiliation{Zhejiang Province Key Laboratory of Quantum Technology and Device, Department of Physics, Zhejiang University, Hangzhou 310027, P. R. China}

\author{Yunhao Lu}
\affiliation{Zhejiang Province Key Laboratory of Quantum Technology and Device, Department of Physics, Zhejiang University, Hangzhou 310027, P. R. China}

\author{Mianzeng Zhong}
\affiliation{School of Physics and Electronics, Hunan Key Laboratory for Super-microstructure and Ultrafast Process, Central South University, Changsha 410083, P. R. China}

\author{Kenji Watanabe}
\affiliation{National Institute for Materials Science, 1-1 Namiki, Tsukuba, 305-0044, Japan}

\author{Takashi Taniguchi}
\affiliation{National Institute for Materials Science, 1-1 Namiki, Tsukuba, 305-0044, Japan}

\author{Qinling Xia}
\email{qlxia@csu.edu.cn}
\affiliation{School of Physics and Electronics, Hunan Key Laboratory for Super-microstructure and Ultrafast Process, Central South University, Changsha 410083, P. R. China}

\author{Zhu-An Xu}
\email{Zhuan@zju.edu.cn}
\affiliation{Zhejiang Province Key Laboratory of Quantum Technology and Device, Department of Physics, Zhejiang University, Hangzhou 310027, P. R. China}
\affiliation{Collaborative Innovation Centre of Advanced Microstructures, Nanjing University, Nanjing 210093, P. R. China}

\author{Yi Zheng}
\email{Correspondence and requests for materials should be addressed to Y.Z. (email: phyzhengyi@zju.edu.cn)}
\affiliation{Zhejiang Province Key Laboratory of Quantum Technology and Device, Department of Physics, Zhejiang University, Hangzhou 310027, P. R. China}
\affiliation{Collaborative Innovation Centre of Advanced Microstructures, Nanjing University, Nanjing 210093, P. R. China}

\date{\today}

\begin{abstract}

Exciting phenomena may emerge in non-centrosymmetric two-dimensional (2D) electronic systems when spin-orbit coupling (SOC) interplays dynamically with Coulomb interactions, band topology, and external modulating forces, etc \cite{science-Skyrmions-Tokura,TbMnO3-nature,QSH-HgTe_ZhangSC_science_06,Rashba-SOC-NM-15,Rashba-SOC-NM-15,Niu-TMDC-QHE-PRL-13}. Here, we report illuminating synergetic effects between SOC and Stark in centrosymmetric few-layer black arsenic (BAs), manifested as giant Rashba valley splitting and exotic quantum Hall states (QHS) reversibly controlled by electrostatic gating. The unusual finding is rooted in the puckering square lattice of BAs, in which heavy $4p$ orbitals form highly asymmetric $\Gamma$ valley with the $p_{z}$ symmetry and $D$ valleys of the $p_{x}$ origin, located at the Brillouin zone (BZ) center and near the time reversal invariant momenta of $X$, respectively. When the structure inversion symmetry is broken by perpendicular electric field, giant Rashba SOC is activated for the $p_{x}$ bands to produce strong spin-polarized $D^{+}$ and $D^{-}$ valleys related by time-reversal symmetry, coexisting with weak $\Gamma$ Rashba bands constrained by the $p_{z}$ symmetry. Intriguingly, strong Stark effect shows the same $p_{x}$-orbital selectiveness for $D$, collectively shifting the valence band maximum of $D^{\pm}$ valleys to exceed the $\Gamma$ pockets. Such an orchestrating effect between SOC and Stark allows us to realize gate-tunable spin valley manipulations for 2D hole gas, as revealed by unconventional magnetic field triggered \textit{even-to-odd} transitions in QHS. For electron doping, the quantization of the $\Gamma$ Rashba bands is characterized by peculiar density-dependent transitions in band topology from two parabolic valleys to a unique inner-outer helical structure when charge carrier concentrations increase. 

\end{abstract}

\maketitle





The coupling of electron spin and orbit degrees of freedom is one of the most profound effects in crystals, playing as the foundation of various fascinating phenomena, such as antisymmetric Dzyaloshinskii-Moriya exchange interaction \cite{science-Skyrmions-Tokura}, magnetoelectric coupling \cite{TbMnO3-nature}, topological quantum states \cite{QSH-HgTe_ZhangSC_science_06}, and the celebrated Dresselhaus and Rashba splitting in energy bands \cite{Rashba-SOC-NM-15}. To activate such spin-orbit coupling (SOC), a lack of spatial inversion center is a necessity, by breaking either bulk- or structure inversion symmetry of crystal lattices. The emergence of van der Waals (vdW) crystals in recent years opens unprecedented opportunities in exploring SOC physics in the two-dimensional (2D) limit \cite{Niu-TMDC-QHE-PRL-13,TMDC-SOC-PRX-16}, when the coupling is tuned by atomic-precision control of quantum confinement and by external modulation of electric field, charge doping and/or vdW heterostructure. 

\begin{figure*}
\includegraphics[width=7 in]{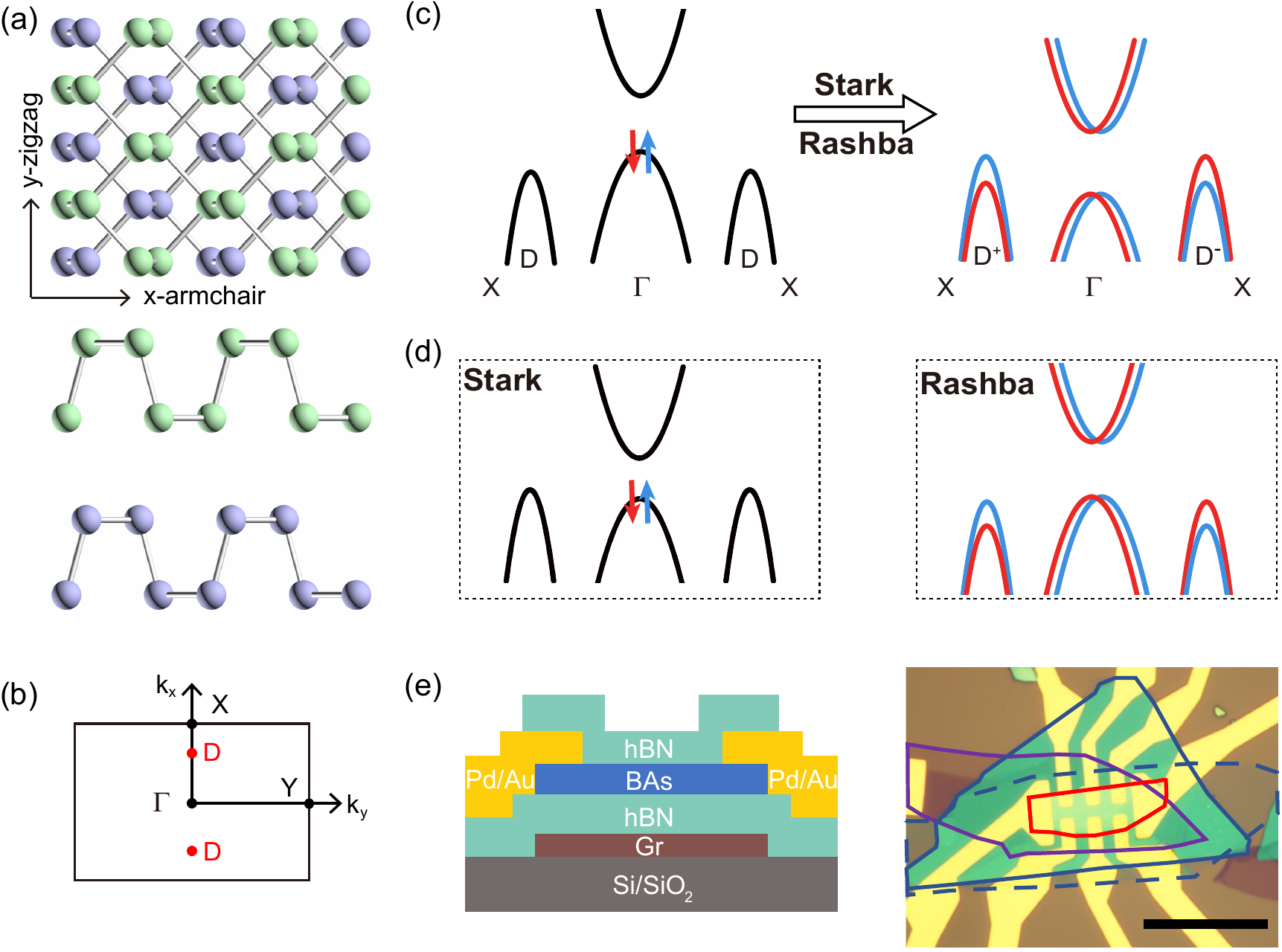}
\caption{\textbf{Gate-tunable giant Rashba valley splitting in BAs by synergetic Rashba and Stack effects.} \textbf{a}, Top and side views of BAs crystal structure. BAs monolayers are AB stacked, while in the vdW plane, the puckering square lattice forms distinctive zigzag and armchair edges. \textbf{b},  The first Brillouin zone of BAs. Unlike BP, the degenerate $D$ valleys in BAs are only 0.2 eV lower in VBM energy than the $\Gamma$ valley. \textbf{c}, Energy bands' evolution of BAs with synergetic Stark effect and Rashba SOC under $E_{z}$, which create spin-polarized $D^{\pm}$ valleys with higher VBM energy than $\Gamma$. \textbf{d}, Energy bands of BAs under Stark effect (left panel) or Rashba SOC (right panel) only. \textbf{e}, Left Panel: Schematic cross section diagram of hBN/BAs/hBN/graphite devices. Right Panel: Optical image of a BAs FET, in which BAs channel and graphite $V_{g}$ are outlined by red and purple lines, while the top and bottom hBN layers are indicated by solid and dashed blue lines respectively. The scale bar is 20 $\mu$m.}
\label{fig1}
\end{figure*}

Here, we report the discovery of gate-tunable giant Rashba valley splitting with spin-valley flavors and unconventional quantum Hall states in few-layer black arsenic, via synergetic SOC and Stark effects with the same orbital geometry selectivity. Black arsenic (BAs) is the group-V analogue of the renowned black phosphorus (BP)  \cite{ZhangYB_14BPFET_NatNano, Xia-BP-NatRevPhy-19}, sharing the same centrosymmetric square lattice in which puckered monolayers are AB-stacked along the $c$ axis (Figure \ref{fig1}a). The driven force to form such a puckered square lattice, symbolized by highly anisotropic \textit{zigzag} and \textit{armchair} crystallographic axes, is resonant bonding of orthogonal $p$ orbitals to satisfy the condition of half-filled bands \cite{ResonantBonding_NM08, ResonantBonding_Wuttig_AM19}. The resulting semiconductors have rather unique electronic structure with $p$-orbitals dominating both valence band (VB) and conduction band (CB), showing multiple valleys with distinct $p$-orbital symmetry (Fig. \ref{fig1}b). Explicitly, for all BP-type materials, the VB maximum (VBM) and CB minimum (CBM) at the Brillouin zone center (designated as $\Gamma$ valley) are predominated by $p_{z}$ orbitals, while $p_{x}$ orbitals form another set of band extremums along the $\Gamma-X$ directions, \textit{i.e.} doubly degenerate $D$ valleys enforced by inversion symmetry (IS) and time reversal symmetry (TRS). Intriguingly, the relative VBM energy between $\Gamma$ and $D$ valleys are determined by the strength of $sp$-hybridization. For BP, substantial $3s3p$ hybridization pushes the anti-bonding VBM of the $\Gamma$ valleys far above the $D$-valley top, producing the well-known zone-centered direct band gap. For BAs, $4s4p$ hybridization is much weaker than BP, resulting in very close VBM energy between $D$ and $\Gamma$ valleys ($\sim 0.2$ eV), as demonstrated by angle-resolved photoemission experiments \cite{BAs-Anisotropy-AM-18}. Note that for binary BP of SnSe, $sp$-hybridization is negligible and the $\Gamma$-valley VBM is well below the $D$ valleys, which form an indirect band gap with the $\Gamma$-valley CBM \cite{SnSe_SdH_NatCommun}.





We now elucidate the synergetic effects between SOC and Stark with the same $p_{x}$-orbital selectivity, which is the key to realize reversible control of spin-valley flavored Rashba valley splitting in BAs (Fig. \ref{fig1}c). It is known that the quantization of few-layer BP in high magnetic field above 30 T is prevailed by Zeeman effect \cite{ZhangYB_16BPQHE_NatNano}, while SOC is negligible due to light atomic weight of phosphorus element. One period larger than P, As has considerable SOC of $0.43$ eV \cite{SOC-Winkler03}, which can induce large energy band splitting when out-of-plane electric field breaks the structure inversion symmetry, \textit{i.e.} extrinsic Rashba effect. Qualitatively, such external electric field induced band splitting can be described by,
\begin{equation}
\hat{H}_{R}= \frac{\alpha_{R}}{\hbar} \bm{\sigma} \cdot (\bm{E_{z} \times p})
\end{equation}
 in which $\alpha_{R}$, $\bm{\sigma}$, $\bm{E}_{z}$, and $\bm{p}$ are Rashba parameter, Pauli spin matrices, perpendicular electric field, and momentum, respectively. From this formula, we can already see the orbital selectiveness of $\bm{E}_{z}$-induced SOC band splitting, which is small for the $\Gamma$ valley due to $p_{z}$ symmetry but becomes maximized for the $D$ bands with large $p_{x}$ momentums in the vicinity of time reversal invariant momenta (TRIM) of $X$. By activating Rashba SOC, the $D$ valleys become spin-valley polarized as designated by $\pm$ flavors, representing the product of spin ($\sigma_{z}$) and valley  ($\tau_{z}$) indexes for the top spin channel of each $D$ valley respectively (Fig. \ref{fig1}c). Such flavor-dependent Rashba band splitting of $D^{\pm}$ is centering on TRIM points of $X$, distinct from the well-known $\Gamma$-centered Rashba effect in BiTeI \cite{Rashba-BiTeI-13,Rashba-SOC-NM-15,Rashba-BiTeI-APRES-NM-11}.



On the other hand, BP shows giant Stark effect (GSE) \cite{Louie04_PRB,Lv-BP-Stark-17}, which can even completely close the band gap by heavy electron doping \cite{BP-Dirac-ARPES-15}. External electric field also induces GSE in BAs, where a profound orbital selectivity for $p_{x}$-bands emerges to push the VBM of the $D$ valleys above $\Gamma$ (Fig. \ref{fig1}d). The different response of the $\Gamma$ and $D^{\pm}$ valleys to $\bm{E}_{z}$ represent a larger GSE coefficient for the latter [see Note 1 in Supplementary Information (SI) for detailed discussions]. Similar orbital-dependent GSE has been systematically explored for transition metal dichalcogenides (TMDCs), which reveal that $\bm{E}_{z}$ weakens interlayer coupling by depleting $p_{z}$ orbitals and redistributing charges into the vdW planes \cite{Towe11_PRB_Stark}. The collective $p_{x}$-orbital selectivity for SOC and Stark effects causes drastic VBM reconstruction when the spin-polarized $D^{\pm}$ valleys gradually take over the $\Gamma$ valley as $\bm{E}_{z}$ increases, making the reversible control of giant flavor-dependent Rashba valley splitting in BAs feasible (see schematic in Fig. \ref{fig1}c). Using temperature($T$)- and angle($\theta$)-dependent magneto transport, we probe the quantum signatures of such reversible Rashba valley splitting, which manifests as exotic \textit{even-to-odd} transitions in quantum Hall states when Fermi level ($E_{F}$) is crossing two sets of unconventional spin-valley asymmetric Landau levels and moving towards the massive Dirac points of the $D^{\pm}$ valleys.

Figure \ref{fig1}e shows the typical device structure of BAs thin-film field-effect transistors (FETs).  BAs flakes were micro-mechanically exfoliated from natural single crystals \cite{Arsenolamprite-1970, BAs-Anisotropy-AM-18, BAs-FET-AFM-18} in a glovebox with argon atmosphere. BAs layer thickness ($10-25$ nm) were estimated by optical contrast and ultimately determined by non-contact atomic force microscopy (see SI Note 2). Compared to BP, BAs micro-flakes exhibit impressive air stability, as evident by robust and repeatable Raman spectroscopy over tens of hours of ambient exposure. Nevertheless, for quantum transport measurements, device protection and isolation from the underlying SiO$_2$ substrate are decisive to observe the intrinsic quantization behavior of BAs. To achieve this, we encapsulated BAs FETs between two hexagonal boron nitride (hBN) layers while an extra graphite layer was lying at the bottom to serve as the back gate ($V_{g}$), a device structure used for studying quantum Hall effects (QHE) in BP \cite{ZhangYB_16BPQHE_NatNano, ZhangYB_18QHE_ntype}. The hBN flakes effectively isolate BAs FET channels from ambient exposure, and the bottom hBN ($\sim20$ nm) is slightly thicker than the top one ($\sim 5$ nm) since the former also serves as the dielectric layer for the back gate ($\sim 3-5$ nm). 

\begin{figure*}
\includegraphics[width=7 in]{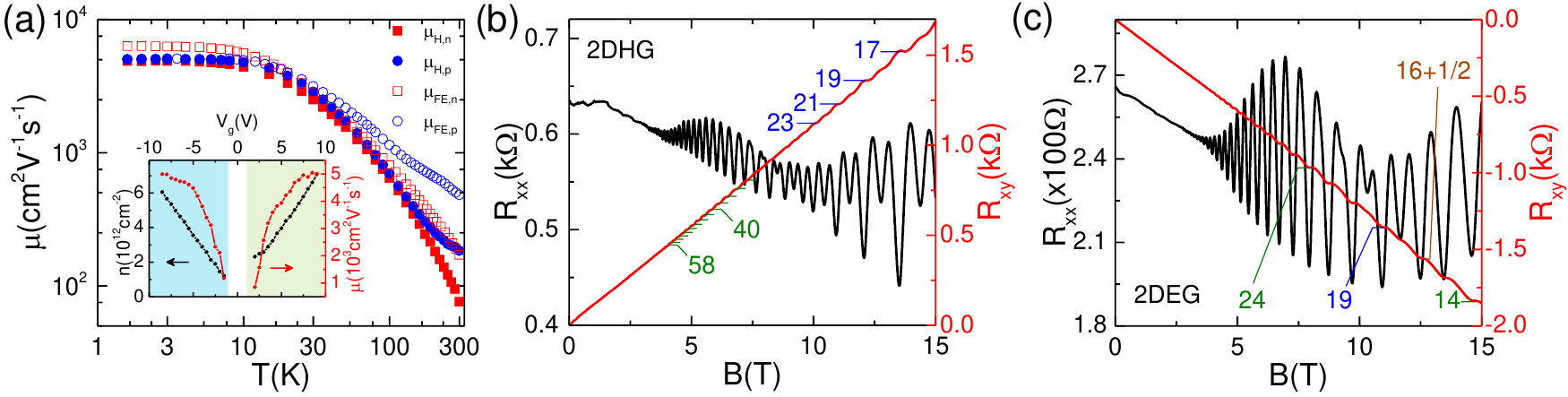}
\caption{\textbf{SdH oscillations and quantum Hall states transitions in BAs 2DHG and 2DEG.} \textbf{a}, $T$-dependent Hall mobility $\mu_{H}$ measured with $V_{g}=+7$ V (electron) and $V_{g}=-7$ V (hole). Inset: Ambipolar FET transfer characteristics at 0.27 K, demonstrating $V_{g}$-controlled Fermi level shifting from CB to VB. \textbf{b}, \textit{Even-to-odd} QHS transitions in BAs 2DHG with 5.30$\times$10$^{12}$ cm$^{-2}$ hole doping. \textbf{c}, SdH oscillations in BAs 2DEG with 5.03$\times$10$^{12}$ cm$^{-2}$ electron doping. }
\label{fig2}
\end{figure*}

The high quality of our BAs devices were evident by excellent field effect transfer characteristics (see Figure \ref{fig2}a and SI
Note 3). As a function of $V_{g}$, BAs FETs show ambipolar transitions from electron to hole conduction for positive and negative gate bias respectively (inset of Figure \ref{fig2}a). Using Pd/Au contacts, we can readily achieve Hall mobility ($\mu_{H}$) exceeding 5000 cm$^{2}$V$^{-1}$s$^{-1}$ for both 2D electron gas (2DEG) and hole gas (2DHG) at 1.6 K, which is truly remarkable and comparable to the best quality BP 2DHG. In the following, we show consistent results from two representative BAs FETs, with rather different hole $\mu_{H}$ of 1600 (BAs-S2) and 5100 cm$^{2}$V$^{-1}$s$^{-1}$ (BAs-S9) respectively. Similar to BP, BAs exhibits $T$-dependent $\mu_{H}$, which becomes saturating below 30 K (Figure \ref{fig2}a). Typically, when $T$ is lowered from 300 K to 1.6 K, $\mu_H$ increases from about 100 cm$^{2}$V$^{-1}$s$^{-1}$ to 5000 cm$^{2}$V$^{-1}$s$^{-1}$ for both hole and electron doping. Figure \ref{fig2}a also demonstrates the wide tunability of carrier densities in the range of $1-8\times10^{12}$ cm$^{-2}$, which is linearly proportional to $V_{g}$ at $T = 0.27$ K. The increase in $\mu_H$ in response to carrier density accumulation suggests that disorder scattering correlated to residual impurities in BAs can be effectively screened by mobile carriers, in agreement with high quality BP devices \cite{ZhangYB_16BPQHE_NatNano,WangN_16_UltraHigh}.




With the extraordinary carrier mobility of our BAs devices, we are able to observe fascinating Shubnikov–de Haas (SdH) oscillations in the magnetoresistance $R_{xx}$ and exotic quantum Hall states (QHS) transitions in the Hall resistance $R_{xy}$ for gate-tunable Rashba valley formation with unusual flavor dependence, the very first report among 2D materials to the best of our knowledge. As shown in Fig. \ref{fig2}b and \ref{fig2}c for the ambipolar BAs-S9 FET, robust SdH peaks emerge at low magnetic field ($B$) below 2 T for both 2DEG and 2DHG at 0.27 K. Using the empirical formula $\mu \sim 1/B_{c}$, we deduce an equivalent carrier mobility of $\sim 5000$ cm$^{2}$V$^{-1}$s$^{-1}$, in good agreement with the Hall results. Much to our surprise, well-defined QHS with an astonishing maximum filling factor ($\nu$) of 58 develop in $R_{xy}$ for 2DHG of $5.30\times 10^{12}$ cm$^{-2}$, accompanying SdH minima precisely. By plotting $R_{xx}$ and $R_{xy}$ together, it is conspicuous to see a pronounced field-dependent transition in the SdH oscillations from even-$\nu$ QHS sequences to odd-$\nu$ QHS plateaus. However, with a lower hole doping of $2.94\times 10^{12}$ cm$^{-2}$, SdH oscillations of 2DHG start with odd-$\nu$ QHS directly (see SI Note 4 and 5 for more details). For both cases, Zeeman splitting induced double peaks become distinctive features in $R_{xx}$ for $B$ above 6 T, where the local double-peak minima correspond to kinks instead of well-developed plateaus in $R_{xy}$. In stark contrast, SdH oscillations of 2DEG have drastically different quantization behavior. As shown in Fig. \ref{fig2}c, the QHS plateaus of $n=5.03\times 10^{12}$ cm$^{-2}$ 2DEG are alternatingly even, odd and half-integer. As we will clarify in the subsequent paragraphs, the contrasting QHS behavior between 2DHG and 2DEG represent the quantum signatures of rather different band topology between the VB and CB of BAs under $\bm{E}_{z}$, which split into flavor-dependent $D^{\pm}$ Rashba valleys and conventional $\Gamma$-centered Rashba bands, respectively. 

\begin{figure*}
\includegraphics[width=7 in]{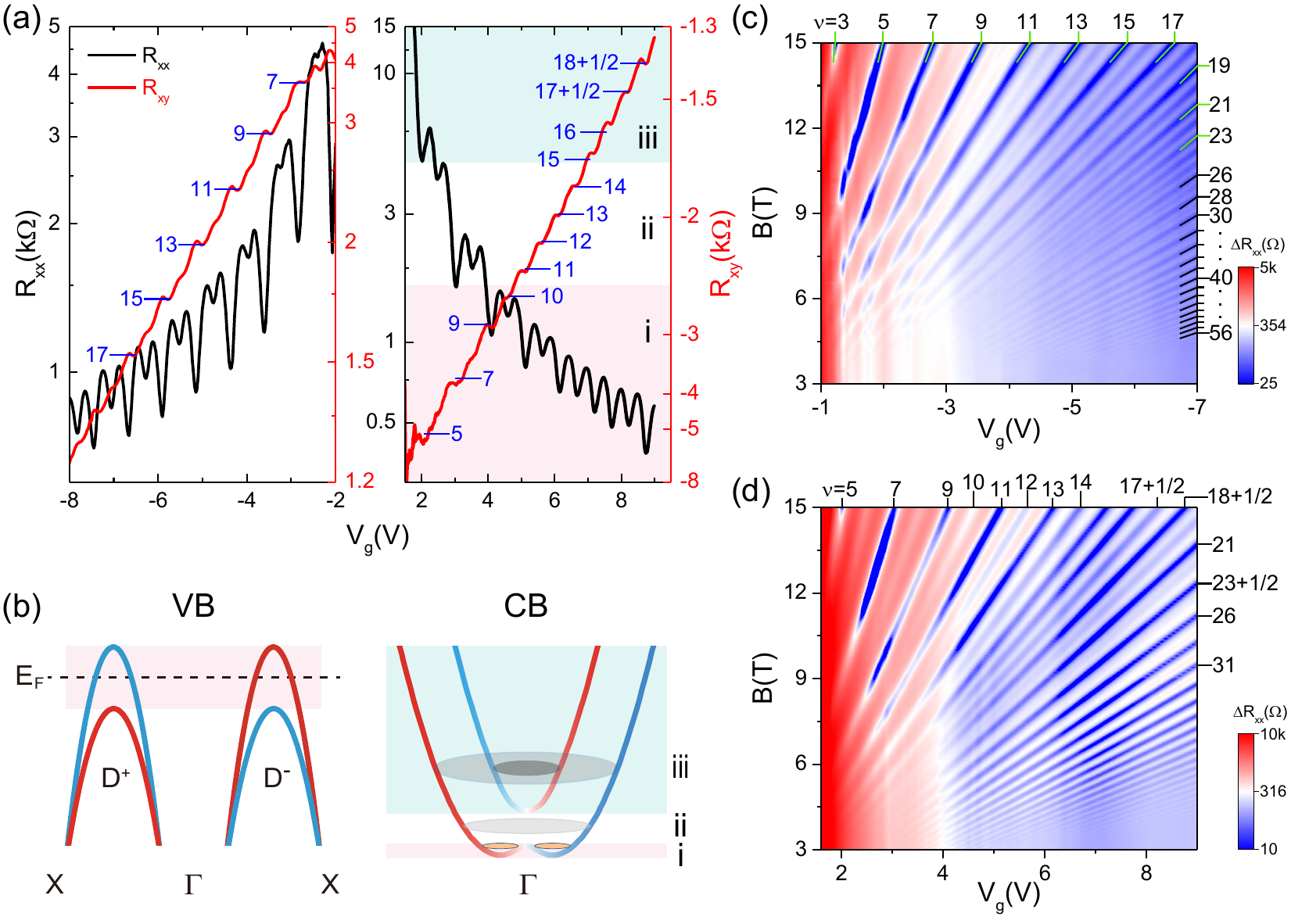}
\caption{\textbf{Unconventional quantum Hall signatures of gate-tunable Rashba valley formation in BAs 2DHG and 2DEG.} \textbf{a}, $R_{xx}$ and $R_{xy}$ as a function of $V_{g}$ at $B=$15 T and $T=0.27$ K, unfolding all odd-$\nu$ QHS sequence for 2DHG but density-dependent transitions in $\nu$ from odd, integer to half-integer for 2DEG. \textbf{b}, Rashba band formation of the VB and CB of BAs under external electric field. The former forms TRS-paired $D^{\pm}$ valleys like monolayer TMDCs, while the latter is the well-known $\Gamma$-Rashba bands. Note that the $\Gamma$-Rashba splitting is exaggerated for clarity. \textbf{c, d} 2D mapping of $R_{xx}$ versus $B$ and $V_g$ for 2DHG and 2DEG, respectively. The mapping data were collected by $V_{g}$ sweeping with fixed $B$, which was increased by steps of 0.1 T for $B\geqslant 9$ T and 0.05 T for $B<9$ T.}
\label{fig3}
\end{figure*}

To get insights into the exotic QHS transitions in 2DHG and 2DEG, we have measured $R_{xx}$ and $R_{xy}$ as a function of carrier doping by continuously tuning $V_{g}$ across the band gap, while keeping the devices at 0.27 K and in constant $B$ of 15 T. As shown in Figure \ref{fig3}a, for 2DHG, odd-$\nu$ [$\nu=2n+1$, where $n$ is the Landau (LL) index] QHS are predominant over the whole doping range, while the lifting of spin degeneracy by magnetic field leads to the formation of Zeeman double peaks centering discernible even-$\nu$ ($\nu=2n$) kinks in $R_{xy}$. Using these experimental data, we deduce the ratio between cyclotron energy ($E_{C}=\hbar\, eB/m^{\ast}$, in which $\hbar$ and $m^{\ast}$ represent the reduced Plank constant and effective mass respectively) and Zeeman spin-splitting energy ($E_{S}=g^{\ast}\mu_{B}B$, where $g^{\ast}$ and $\mu_{B}$ are the effective Land$\acute{e}$ g-factor and the Bohr magneton respectively) to be $E_{S}\approx 0.49 E_{C}$ (see SI Note 6). Unlike MoS$_{2}$ and WSe$_{2}$, both show prominent density dependence in $g^{\ast}$ as a result of enhanced electron-electron interaction with lowering doping, $E_{S}$ in BAs decreases slightly when hole doping changes from $5.66\times10^{12}$ cm$^{-2}$ ($V_{g}=-8$ V) to $2.13\times10^{12}$ cm$^{-2}$ ($V_{g}=-3$ V), which allows us to exclude the possible origin of the aforementioned \textit{even-to-odd} QHS transitions of 2DHG in the competition of $E_{C}$ and $E_{S}$. Such a claim is also strongly supported by $\theta$-dependent $R_{xx}$ in rotating magnetic field \cite{WSe2-Gamma-Zeeman-PRL-17}, which reveal no significant shifting in Zeeman peaks when $B$ is rotated from 0$^{\circ}$ to above 60$^{\circ}$ (See SI Note 6). 

Using positive $V_{g}$, the density-dependent QHS of 2DEG reveal peculiar transitions in band topology when $E_{F}$ is tuned in the CB. As shown in Fig. \ref{fig3}a, for $V_{g}<5$ V, $n$-type QHS exhibit the same odd-$\nu$ ($2n+1$) main sequences, while the local $R_{xx}$ valleys of Zeeman double peaks correspond to weaker even $\nu$. For $V_{g}>5$ V, QHS are robust for both even and odd filling factors, as evident by $\nu=11$, 12, 13, and 14 plateaus. Even more strikingly, QHS become half-integer quantized in the form of $\nu^{\prime}=\nu+1/2$, for $\nu=16$, 17, and 18 plateaus. These exotic changes in 2DEG QHS are rooted in the formation of two $\Gamma$-centered Rashba bands in the CB, when spin degeneracy are lifted by $\bm{E}_{z}$-activated SOC. Unlike the protected gapless surface states in topological insulators, the $\Gamma$-Rashba band crossing point in BAs is topologically trivial and becomes gapped by external  field. At 15 T and 0.27 K, the gapped $\Gamma$-Rashba bands can be divided into three QHS zones, as sketched in Fig. \ref{fig3}b. The first zone is for $E_{F}$ below the Zeeman gap ($E_{S}$), corresponding to low positive $V_{g}$ when the Fermi surface (FS) is characterized by two elliptical pockets produced by weak Rashba splitting and strong in-plane band anisotropy. When quantized, the LLs of these two isolated Fermi pockets are equivalent to the celebrated flavor-dependent LL spectrum of heavily massive Dirac fermions in MoS$_{2}$, in which QHS plateaus in the vicinity of valley extrema follow $\nu=2n+1$ \cite{Niu-TMDC-QHE-PRL-13}. On the other hand, when $E_{F}$ is well above the Zeeman gap for large $V_{g}$, the FSs form unique inner-outer nested structure by intersecting two helical Rashba bands \cite{Rashba-BiTeI-APRES-NM-11,PtBi2_HeSL_NC19}. Due to the helical spin-momentum locking, a $\pi$ Berry phase emerges in QHS manifested by $\nu=n+1/2$, where the half integer is the quantum transport fingerprint of massless fermions in topological materials \cite{Gr05Geim_QHE, Gr05-QHE-KIM, Rashba-BiTeI-13,QHE-BiSbTeSe2_ChenYP_NP14}. In between these two zones, $V_{g}$ pushes $E_{F}$ right into the gap region, creating one large spin-polarized pocket. The LLs of such a non-degenerate band are characterized by integer $\nu$.


The relation between charge carrier polarity and QHS transitions become more intuitive by 2D mapping of $R_{xx}$ as a function of both $V_{g}$ and $B$, in which well-defined even and odd QHS states for different SdH minima are labelled by solid lines and $\nu$ indexes (Fig. \ref{fig3}c and \ref{fig3}d). Note that for 2DEG, the horizontal indexing of $\nu$ is referring to 15 T, while the vertical $\nu$ is determined by SdH oscillations at $V_{g}=9$ V. For 2DHG, the \textit{even-to-odd} transitions are located at $\nu=24$, which defines the lower boundary of odd-$\nu$ QHS in Fig. \ref{fig3}c. In contrast, odd-$\nu$ QHS are both field- and doping-dependent, and thus distributed extensively over the 2D $R_{xx}$ image of 2DEG (Fig. \ref{fig3}d). The different QHS behavior between 2DHG and 2DEG can not be simply attributed to negligible band asymmetry between the CBM and VBM of the $\Gamma$ valley (see SI Note 1). Instead, it is the manifestation of collective $p_{x}$-orbital selectivity for Rashba SOC and Stark effects, which reverse the VBM energy ordering between the $\Gamma$ and $D^{\pm}$ valleys when negative $V_{g}$ is applied. As depicted in Fig. \ref{fig1}d, the Rashba energy splitting of the $D$ valleys resembles Zeeman effect, rather distinct from the established $\Gamma$-Rashba effect. This is because the momentum-dependent band splitting of the $D$ valleys is referenced to the TRIM points of $X$, which are local band minima of the top VB band. When viewed in the first BZ, the Rashba-split $D^{\pm}$ are spin-valley polarized as required by TRS. 

The quantization of the spin-valley flavored $D^{\pm}$ valleys shares exactly the same LL structure as  monolayer TMDCs, when Rashba SOC creates two groups of LLs (designated by $n_\mathrm{I}$ and $n_\mathrm{II}$ repectively) by breaking both spin and valley degeneracy. As sketched in Fig. \ref{fig3}b, the classification of LLs for the $D^{\pm}$-Rashba valleys are based on \textit{flavors}, \textit{ i.e.} the sign of $\tau_{z} \sigma_{z}$ ($\pm 1$). TRS between the $D^{+}$ and $D^{-}$ valleys enforces double degeneracy in each set of LLs, with the only exception of spin lifted $n_{i}=0$ LLs \cite{Niu-TMDC-QHE-PRL-13}. For low negative $V_{g}$, $E_{F}$ only intersects $n_\mathrm{I}$-group LLs, and $\nu$ follows the odd sequences of $2n+1$ when $B$ linearly increases LL degeneracy. By increasing $V_{g}$, the FS will include both groups of LLs in low field, producing even-$\nu$ QHS which will eventually evolve into odd-$\nu$ when $E_{F}$ crosses over $n_\mathrm{II}=0$ LL \cite{Niu-TMDC-QHE-PRL-13}. Note that the spin-up and spin-down Rashba bands of each $D^{\pm}$ valley are further subjected to Zeeman splitting, which also contributes to the dynamic \textit{even-to-odd} QHS transitions for 2DHG.

%



\begin{figure}
\includegraphics[width=3.15 in]{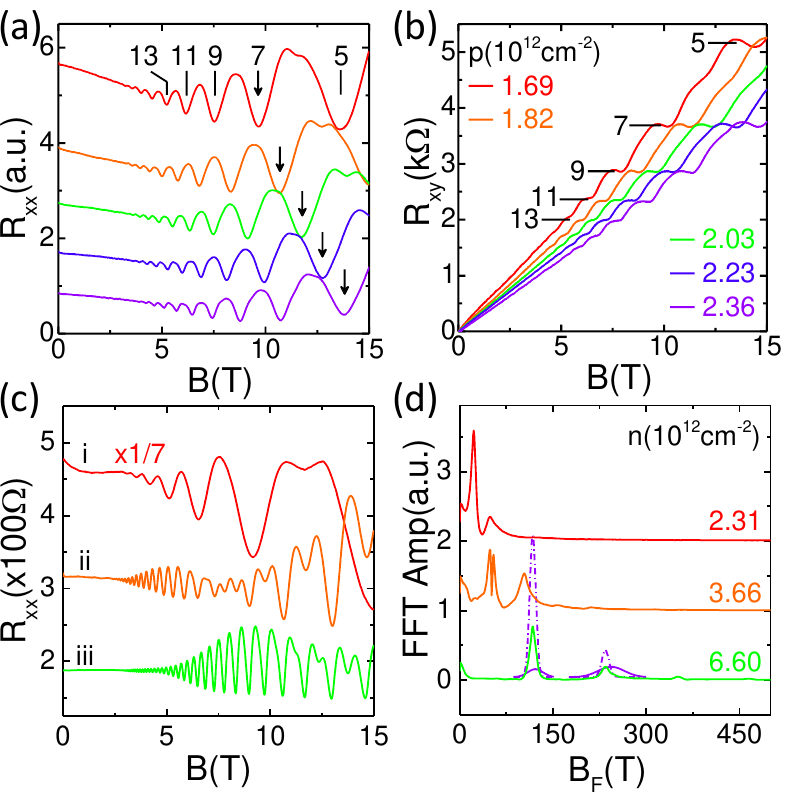}
\caption{\textbf{Rashba band splitting controlled SdH beating patterns in BAs 2DHG and 2DEG.} \textbf{a}, In contrast to high hole doping in Fig. \ref{fig2}b, low-doping SdH beating of 2DHG at 0.27 K shows large periodicity and weak FFT frequency splitting. The black arrows indicate the evolution of $\nu=7$ SdH minima vs hole doping. \textbf{b}, QHS plateaus correspond to \textbf{a}, showing all odd-$\nu$ sequence for low hole doping. \textbf{c}, Density-dependent beating patterns in SdH oscillations of 2DEG, which can be divided into three regimes as defined in Fig. \ref{fig2}b.  (\textbf{d}) The corresponding FFT analyses of \textbf{c}, revealing the correlations between FS topology and SdH beating behavior. The superimposed black lines for $6.60\times 10^{12}$ cm$^{-2}$ represent two Gaussian peaks for fitting the beating frequency. Note that the outer-pocket FFT amplitude is multiplied by 3 for clarify. }
\label{fig4}
\end{figure} 

It is apparent in Fig. \ref{fig2}b and \ref{fig2}c that the emergence of Rashba-splitting bands in both 2DHG and 2DEG for high doping levels ($> 5 \times 10^{12}\, \mathrm{cm}^{-2}$) are coincident with strong beating patterns in SdH oscillations \cite{Gate-SOC-InGaAs-97}, due to the formation of dual inner-outer FSs with close cross-section areas. However, for lower doping, the $\Gamma$- and $D$-valley associated Rashba bands have dissimilar beating behavior, as a result of rather different FS topology illustrated in Fig. \ref{fig3}b. Typical low-doping SdH beating patterns of 2DHG (BAs-S9) are characterized by large periodicity and weak oscillation frequency splitting after fast Fourier transform (FFT), in consistent with two well-separate TRS-valley pockets (Fig. \ref{fig4}a and SI Note 4). Because $E_{F}$ intersects one spin-polarized band for each $D$ valley, the corresponding $R_{xy}$ only shows odd-$\nu$ QHS plateaus (Fig. \ref{fig4}b). For 2DEG, the beating patterns are more dynamically evolved as a function of both doping levels and magnetic field (Fig. \ref{fig4}c and \ref{fig4}d). In the vicinity of CBM (Regime-i in Fig. \ref{fig3}b), the SdH beating of 2DEG is also controlled by two spin-polarized $\Gamma$-Rashba pockets. By increasing doping and pushing $E_{F}$ to the $\Gamma$ Dirac point (Regime-ii in Fig. \ref{fig3}b), the FSs become inner-outer nested, producing strong SdH beating and oscillation frequency splitting. Well above the $E_{S}\, (B=15\,\mathrm{T})$ gap, two helical Dirac FSs are responsible for strong SdH beating with large periodicity (Regime-iii in Fig. \ref{fig3}b).       

The \textit{even-to-odd} transition in QHS of BAs 2DHG would be the first experimental demonstration of unconventional quantum Hall effect of a spin-valley flavored electronic structure \cite{Niu-TMDC-QHE-PRL-13}, which is predicted in monolayer TMDCs but experimentally becomes overshadowed by field- and density-dependent enormous Zeeman effect. The $V_{g}$-driven exotic QHS in 2DEG also show the great potential of BAs in studying unprecedented gate-tunable quantization phenomena. Considering that $B$ is only 15 T in our experiments, we are expecting to see more fascinating SOC and Rashba related physics in higher magnetic field, such as quantum valley Hall and fractional Quantum Hall effect. 




\section*{Methods}

\textbf{BAs FET fabrications.} During device fabrications, BAs/hBN/graphite heterostructures were first assembled by the polycarbonate (PC)-based dry transfer technique \cite{Transfer-PPC,Transfer-PC}, which was followed by preparing Au/Pd contacts (60 nm and 2 nm respectively) to BAs flakes using standard electron beam lithography. Finally, the top hBN layer was transferred on the top of BAs FETs to finish the encapsulation. For micro-mechanical exfoliation, heavily doped silicon substrates with 285 nm SiO$_2$ were used for good optical contrasts for all three types of 2D crystals used in this study. 

\textbf{Density functional calculations.} The first principle calculations of BAs were performed using the Vienna \textit{ab initio} package \cite{VASP} with projector-augmented wave (PAW) \cite{PAW} pseudopotential and generalized gradient approximation (GGA). For few-layer BAs, a 30 \AA\, vacuum layer was adopted to avoid the unphysical interaction. Van der Waals interaction was included by the optB88-vdW exchange functional \cite{opt88andoptPBE} for band calculations and optimization. During the geometry optimization, the lattice and atomic positions were allowed to relax until the force per atom was smaller than 0.001 eV/\AA. The energy cutoff was set to 500 eV and the energy convergence criteria was set to $10^{-6}$ eV. A $\Gamma$-centered $9 \times 10 \times1$ $k$-mesh was adopted for all calculations. The effect of spin-orbital coupling (SOC) \cite{SOC} was included self-consistently due to the considerable SOC energy of $0.43$ eV in As. A sawtooth-type periodic potential was applied in perpendicular to the vdW plane to simulate the electric field $\bm{E}_{z}$.

\textbf{Quantum transport measurements.} SdH oscillations and carrier dependent $R_{xx}$ and Hall resistance $R_{xy}$ are measured with two SR830 lock-in amplifiers using a typical excitation current of 100 nA at 13.3 Hz. The gate voltage are supplied by a Keithley 2400. He$_{3}$ temperature quantum transport measurements were performed using an HelioxVL insert in a 15-T IntegraAC system (Oxford Instruments). A TeslatronPT cryofree system (Oxford Instruments) was employed to measure angle-dependent and T-dependent $R_{xx}$ and $R_{xy}$ from $1.6-300$ K.

\begin{acknowledgments}
This work is supported by the National Key R\&D Program of the MOST of China (Grant Nos. 2016YFA0300204, 2017YFA0303002 and 2019YFA0308602), the National Science Foundation of China (Grant Nos. 11790313, 11574264, 11774305 and 61904205), Zhejiang Provincial Natural Science Foundation (Grant No. D19A040001), and the Natural Science Foundation of Hunan Province (Grant No. 2020JJ4677). Q.L.X. acknowledges the funding support from the Open Research Fund of State Key Laboratory of Pulsed Power Laser Technology. Y.Z. acknowledges the funding support from the Fundamental Research Funds for the Central Universities.
\end{acknowledgments}

\section*{Author contributions}
Q.L.X. and Y.Z. initiated and supervised the project.  F.S. fabricated BAs devices and carried out all the measurements, assisted by X.K.S. and J.H. C.Q.H and Y.H.L. did the DFT calculations. K.W. and T.T. prepared high-quality boron nitride single crystals. F.S., C.Q.H., Q.L.X. and Y.Z. analysed the data and wrote the paper with inputs from all authors.

\section*{Competing financial interests:}
The authors declare no competing financial interests.

\end{document}